\documentstyle{elsart}
\input psfig.sty

\begin{document}
\begin{frontmatter}

\title{Development of a Gd Loaded Liquid Scintillator for
Electron Anti-Neutrino Spectroscopy}
\author{Andreas G. Piepke$^1$, S. Wayne Moser$^2$ and Vladimir M. Novikov$^{1,3}$}
\address{$^1$ Department of Physics 161-33, Caltech, Pasadena, 
CA 91125, USA\\
$^2$ Bicron/SGIC Inc., Newbury, OH 44065, USA\\
$^3$ Institute of Nuclear Research, 117312 Moscow, Russia}

\begin{abstract}
{We report on the development and deployment of 11.3 tons of 0.1$\%$ Gd 
loaded liquid
scintillator used in the Palo Verde
reactor neutrino oscillation experiment.
We discuss the chemical composition, properties,
and stability of the scintillator elaborating on the details 
of the scintillator preparation crucial for obtaining a good scintillator 
quality and stability.}
\end{abstract}
\end{frontmatter}

\section{Introduction}

The Palo Verde (Arizona) reactor neutrino experiment is a long baseline
disappearance neutrino oscillation search~\cite{proposal}. Its goal is
to test the $\nu_{\mu} \leftrightarrow \nu_e$
solution of the atmospheric neutrino anomaly~\cite{kamio_at}
utilizing the inverse beta decay
$\overline{\nu}_e + p \rightarrow e^+ + n$ as detection reaction.
The Palo Verde detector is a finely segmented device minimizing
correlated backgrounds by exploiting fast triple coincidences
between detector elements.
The positron energy deposit in coincidence with the annihilation quanta
firing at least two neighboring elements identify an e$^+$. 
A delayed
neutron capture signal completes the anti-neutrino signature.\\
The 11.3 tons of hydrocarbon based liquid scintillator (H:C$\approx$2)
serve simultaneously as target and
detector. The measured kinetic energy of the
$e^+$, being stopped in the liquid scintillator, allows
the reconstruction of the $\overline{\nu}_e$ energy.
As the e$^+$ energy is rather low (1-5 MeV), a high light yield scintillator
is vital for obtaining good energy resolution.\\
The correlated positron-neutron signature of the detection reaction
is a powerful tool for the reduction of random backgrounds.
The Gd loaded scintillator was chosen for the Palo Verde project
because it offers two important advantages over pure hydrocarbon
based scintillator:
\begin{itemize}
\item A large thermal neutron capture cross
section of the isotopes $^{155,157}$Gd (61400 and 255000 barns) shorten
the neutron capture time. For
0.1$\%$ Gd loading (by weight) the neutron capture time is 
$\tau \approx 30\;\; \mu s$, compared to $\tau \approx 180\;\; \mu s$ for capture 
on the proton in unloaded scintillators.
\item A release of a high energy (8 MeV) gamma cascade after thermal 
neutron capture on Gd results in a neutron capture signal well above
the radioactivity backgrounds. 
\end{itemize}
The Palo Verde detector
consists of 66 acrylic tanks ($900\; cm \times 12.7\; cm \times 25.4\; cm$)
containing the liquid scintillator. Acrylic was chosen to 
obtain efficient light collection into the 5'' photomultiplier
attached to either side of the tank by exploiting total reflection.
This choice requires
long term compatibility of the cells with the liquid scintillator and
excludes the use of pure solvents like xylene (dimethylbenzene) or 
pseudocumene (1-2-4 trimethylbenzene, PC). Goal of the project was 
to formulate
a liquid scintillator containing not more than 40$\%$ PC diluted with
mineral oil and dissolve the Gd in 
this mixture in a way compatible with the acrylic. Early developments
of this cocktail were done in collaboration with NE Technology Ltd.
The relatively large tank length places strict requirements on 
the transparency of the scintillator.\\
Below we will
discuss the scintillator formulation chosen to
satisfy these requirements. We will present our
measurements of the scintillator properties and, finally,
comment on the stability of the liquid.\\
The technical details given in this paper might be of interest also
for other projects; for example large (100 tons) liquid scintillation
solar neutrinos detectors, loaded with Gd or Yb,
are being discussed~\cite{ragha}. 

\section{Scintillator Formulation}

To assure compatibility of the scintillator with the acrylic tanks
a mixture of pseudocumene and mineral oil was chosen.
As Gd dissolves in neither of the above liquids it was necessary to
prepare a soluble and acrylic-compatible Gd compound.
The compound prepared was gadolinium 2-ethylhexanoate
[Gd(CH$_3$(CH$_2$)$_3$CH(C$_2$H$_5$)CO$_2$)$_3 \cdot $XH$_2$O]. It can be
synthesized from gadolinium chloride, oxide, or nitrate. Early preparations
in the project were made from the nitrate, since a sufficiently radio-pure
source was available. However, solutions from the nitrate were found to be
unstable (solid-liquid phase separation) when handled, especially when
exposed to air. The oxide, Gd$_2$O$_3$, was converted to the 2-ethylhexanoate
in an aqueous reaction, collected by filtration, and dried. The resultant
compound was purity tested and then dissolved in a scintillation solvent, along
with a primary fluorescent additive, a spectrum shifter, an antioxidant,
and small amounts of two additional solvents to help keep the gadolinium
compound in solution. The emission peaks of the primary fluor and spectrum
shifter were 365 nm and 425 nm, respectively.\\
Pseudocumene was the solvent of choice, for its high
scintillation efficiency, lower solvency toward acrylic, and higher
flash point than xylene. To increase the hydrogen content, and further
decrease the solvent attack on the acrylic tanks, a high purity mineral
oil was added as a final component of the finished scintillator solution.
The solution is optimized at 60$\%$ oil by volume as this is the
minimum necessary to dilute the pseudocumene enough that it doesn't
harm the acrylic tank. A higher concentration of oil than this decreases
the scintillation light yield, since this is a function of pseudocumene
content (an aromatic liquid which inherently scintillates, whereas the
oil is a saturated hydrocarbon and does not). Also, the higher the oil
content, the lower the possible Gd content, as this also is a function
of the pseudocumene solvent content.\\
We found that the fully blended scintillator, contained in 200 l steel
drums did not remain stable (a single phase liquid) during the transport
over ground via truck from Ohio to Arizona. To solve this problem, 
pseudocumene based ``concentrate'' was blended in Ohio and additional
pseudocumene and the mineral oil was blended on site in Arizona. 
This procedure is described below.

\section{Scintillator Blending}

The Gd ``concentrate'' (BC-521C)
as prepared by Bicron at their Ohio factory was shipped to 
Palo Verde in 200 l stainless steel drums. There it was blended
with the mineral oil
(Witco Scintillator Fluid), and 1,2,4-trimethylbenzene
(Koch Chemicals obtained through Bicron Inc.).
The proportions used were 1:1:3 by volume for the Gd concentrate,
PC and mineral oil to arrive at a Gd loading of 0.1$\%$ by mass
for the blended scintillator.
Special care was taken to select hardware being
compatible with the used solvents. As the concentrate and mineral
oil do not mix well (rapid addition of the mineral oil may lead
to irreversible precipitation of the Gd compound) care was taken to 
mix those components in a controlled way.\\
The blending was done in two 200 l stainless steel drums equipped with
air driven stirrers having stainless steel shafts and impellers.
With each drum processing about 170 kg of scintillator; the load
of one acrylic cell.
The liquids were moved by air driven diaphragm pumps (Nylon body and
Teflon diaphragm). Separate pumps were used for the mineral oil,
the two other components and the blended scintillator. All piping 
to come in contact with pure 
Gd concentrate or PC was made from Teflon while
PVDF valves were used. 
For the mineral oil polypropylene tubing was
used. The blending drums were placed on electronic scales for
a precise determination of the scintillator weight.\\
For the blending each component was first pumped into an intermediate
stainless steel vessel placed on an electronic scale to determine
its weight and then drained by gravity into the mixing drum.
First the concentrate and PC were mixed. Then the mineral oil was
added at a rate not exceeding $\sim$3 l/min. During the blending process
the liquid was regularly inspected 
for precipitation of the Gd compound. The
stirrers were operated at a low speed to avoid excessive agitation
of the liquid. The final mixture was stirred
for at least 1 hour. The liquid was then pumped through a 200 nm
Gelman ``HiFLO Sol-Vent DCF'' filter directly into
the acrylic tank. 
The filter was changed after every batch of scintillator.
A 2 l sample was retained for every batch blended.\\
The liquid scintillator could not be bubbled with
nitrogen for removal of dissolved oxygen as this bears the risk to
destabilize the mixture leading to precipitation of the
Gd compound.
The acrylic tanks were flushed with Ar before the filling in
order to remove any residual vapors left over from the bonding and to
displace the air. In addition, all tanks were pressure tested prior
to filling to detect possible leaks.\\
A total of 11343 kg of liquid scintillator was prepared in
this way.

\section{Scintillator Quality}

The following scintillator properties
were routinely tested for at least one sample of blended scintillator for every batch 
of Gd concentrate:
\begin{itemize}
\item Light attenuation length at 440 nm
\item Light yield
\item Gd loading
\end{itemize}
Achieving an acceptable light attenuation length was the
most challenging task during the scintillator development.
Before working with production size samples, discussed below,
we went through four generations of prototypes. In this
prototype phase emphasis was given to the optimization of
the solvent balance and quality of the raw ingredients. While
the light yield and Gd loading were satisfactory from the
beginning, early samples did exhibit problems in their
transparency.\\
In the following we will discuss our data on the scintillator
properties.

\subsection{Attenuation Length}

\begin{figure}[htbp]
\centering
\leavevmode
\mbox{\psfig{file=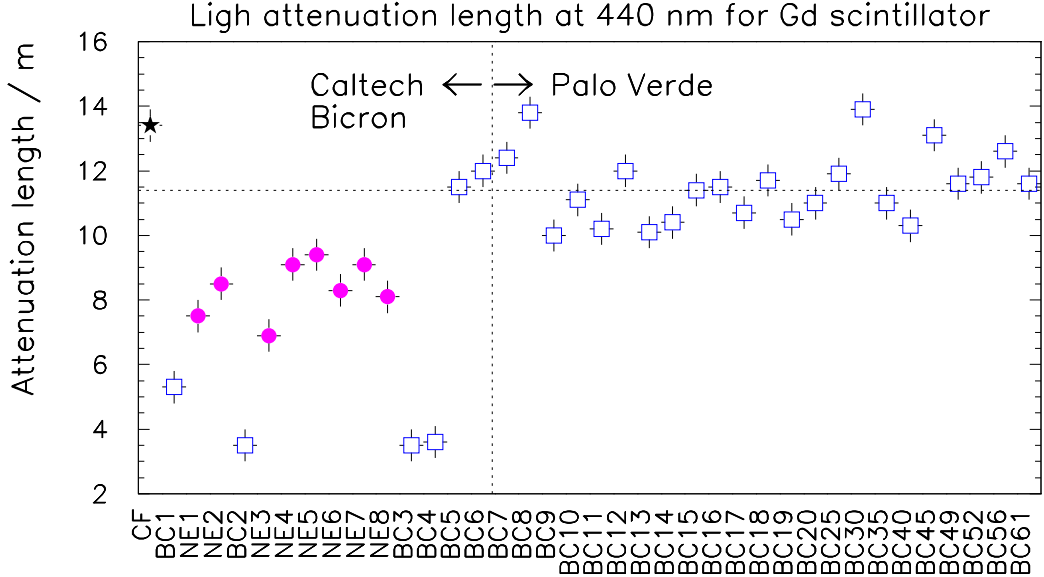,height=10cm,clip=}}
\caption{Light attenuation length at 440 nm as measured for cleaning fluid (CF, star),
NE scintillator (NE, circle) and Bicron scintillator (BC, square). The scintillator
batches are identified by successive numbering, as given in the figure.
All batches to the
right of the vertical line were blended in Palo Verde, while the rest was mixed
at Caltech or Bicron.
}
\label{al}
\end{figure}
The light attenuation length was measured using a vertically oriented
1.5 m long, 3 cm diameter stainless steel tube filled with liquid. 
A blue LED equipped
with a 440 nm wavelength filter and focussed to parallel 
by means of a
7-cm-focal-length lens, placed one focal length away from
a pinhole aperture, illuminates a phototube at the other end
of the tube. The LED was operated in pulsed mode to minimize background
through scintillation light. The average pulse hight, seen by the phototube,
was then measured for varying liquid hight and fitted to
an exponential to obtain the attenuation length (scattering plus absorption).\\ 
Figure~\ref{al} shows the results obtained through those measurements.
The first entry labeled CF denotes a measurement done with a batch of
fluid composed of 36$\%$ PC and 64$\%$ mineral oil mixed in the
hardware described before. The good attenuation length obtained demonstrated that all
used vessels and plumbing had the necessary purity before starting to work with
the scintillator. This liquid was also used to test
one of the acrylic tanks for surface impurities.\\
The batches labeled NE in figure~\ref{al} were early prototypes made
by NE Technology Ltd. The first batches blended by Bicron (labeled BC1-4
in figure~\ref{al}) before transportation showed a rather high light absorption.
The fact that samples BC5 and BC6, blended at Caltech from Gd
concentrate like the NE samples NE1-8, were of much better
transparency shows that the blended 
scintillator is too fragile for a long transport.
After switching to the on-site blending attenuation lengths 
consistently larger than
10 m were achieved for the bulk of the scintillator. For the first
three drums of Gd concentrate, the light attenuation length of
every scintillator batch was tested (BC7-20), for the later samples
only one per concentrate batch. The average attenuation length for
samples blended from BC5 on was 11.4 m.
\begin{figure}[htbp]
\centering
\leavevmode
\mbox{\psfig{file=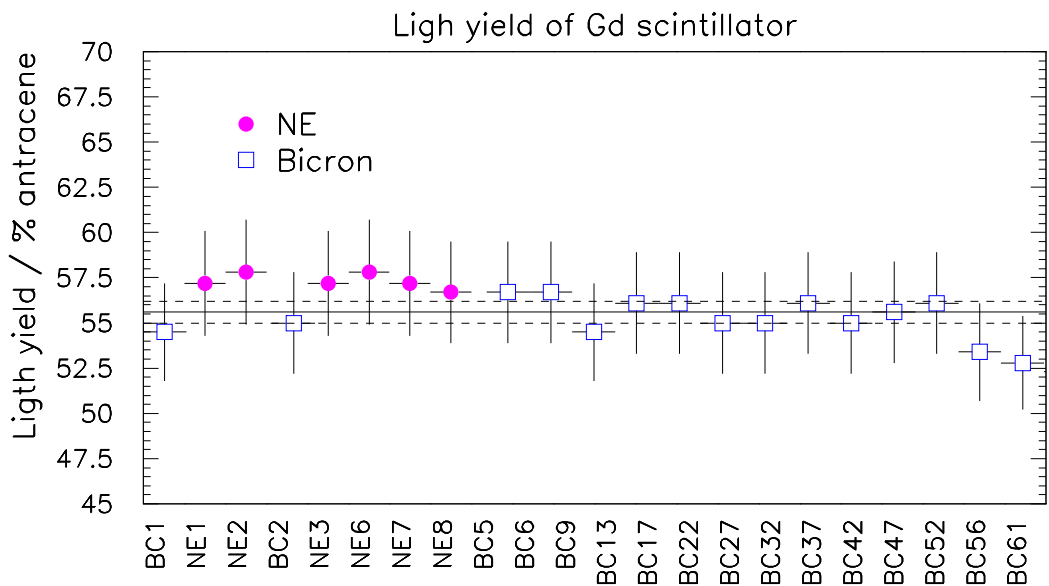,height=10cm,clip=}}
\caption{Light yield of different batches of Gd scintillator. 
}
\label{ly}
\end{figure}
\subsection{Light Yield}

The light yield of the scintillator was measured using a $^{207}$Bi conversion
electron source, irradiating a 2'' diameter Petri dish containing 14 ml
of the sampled scintillator. The resulting liquid hight of 7 mm was chosen
to stop the $\sim$1 MeV conversion electrons (range $\sim$5 mm) within the liquid 
while optimizing the solid angle towards the photomultiplier.
The sample was placed on a 5'' photomultiplier. To monitor PMT stability 
a 1'' diameter NaI detector, separated by a thin Pb shield from the source,
registering its $\gamma$-radiation was placed next to the sample onto the
same PMT and both spectra were collected in parallel. The position of
the $^{207}$Bi conversion electron peak in the measured spectrum is 
a measure of the average pulse hight and hence the light yield. A mixture
of PC with 4 g/l PPO and 100 mg/l bisMSB, deoxiginated by bubbling
with Ar before the measurement, served as a light yield standard. We assume
a light yield of 80$\%$ antracene for this mixture. Figure~\ref{ly} shows
the measured light yield of scintillator samples. The light yield was verified
for at least on scintillator batch per concentrate lot.
The light yield was consistently 56$\%$ antracene for the 22
tested samples.

\subsection{Gd Loading}

The Gd concentration of the liquid was measured using x-ray fluorescence
after thermal neutron capture on Gd, as described in~\cite{novikov}.
The measurements were done in comparison to standard solutions of
Gd(NO$_3$)$_3$ dissolved in alcohol. The absolute Gd concentration of 
one scintillator batch 
was verified by mass spectroscopy in a commercial laboratory.
In addition to this cross check a few early samples were also tested
by neutron activation of $^{158}$Gd, using a $^{252}$Cf spontaneous fission
source placed in a water moderator. The 363 keV $\gamma$-radiation
following the $\beta$-decay of $^{159}$Gd into $^{159}$Tb (T$_{1/2}$=18.6 h) 
was detected with a Ge detector.\\
\begin{figure}[htbp]
\centering
\leavevmode
\mbox{\psfig{file=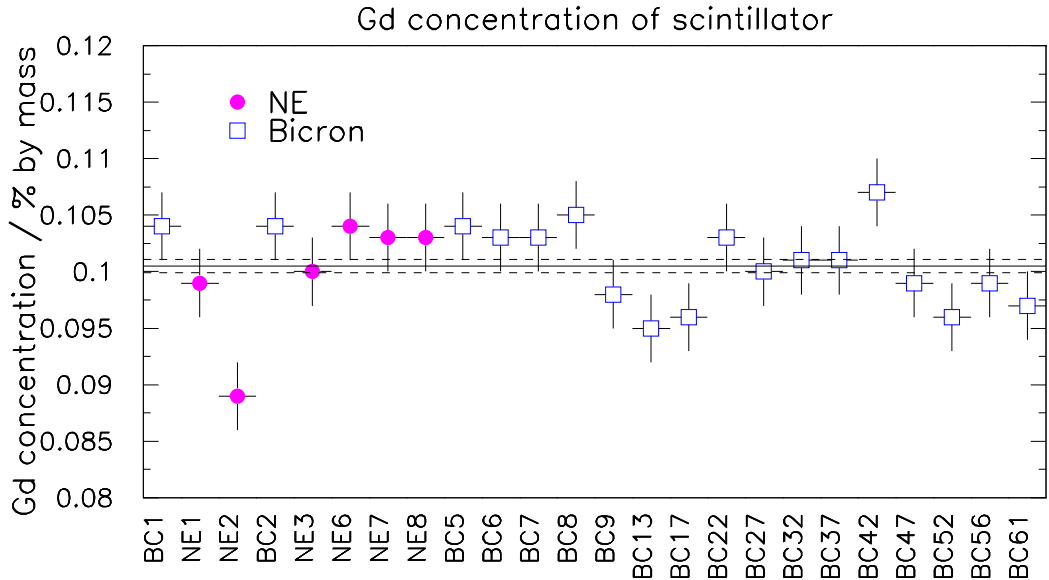,height=10cm,clip=}}
\caption{$\%$ Gd loading (by mass) of scintillator batches.
}
\label{gd}
\end{figure}
As for the light yield at least one scintillator batch per concentrate lot
was tested for its Gd loading to make sure that the detector is of
reasonable uniformity. As can be seen in figure~\ref{gd} the Gd loading
was consistently 0.1$\%$. The small scattering shows that the production
process of the concentrate is well controlled. 
The low value of NE2 was confirmed by the observation
of Gd precipitation in the transport drum. The solvent balance of all following
batches was readjusted after this observation. For prototype samples
kept at Caltech for several years no instability of the Gd loading was observed.
The described loading scheme is quite stable if the liquid is
not disturbed or agitated.

\section{Scintillator Stability}
\begin{figure}[htbp]
\centering
\leavevmode
\mbox{\psfig{file=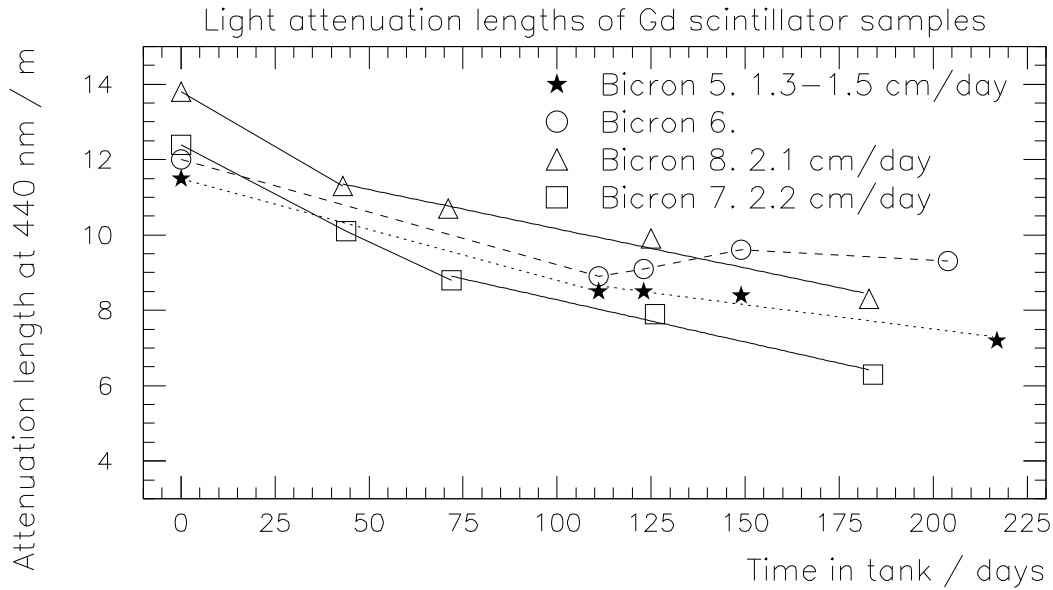,height=10cm,clip=}}
\caption{Time development of the light attenuation length for Gd scintillator batches.
Samples have been taken from the acrylic tank.
}
\label{stab}
\end{figure}
As neither light yield nor Gd loading showed any observable degradation over time
we shall concentrate in the following on the time development of the scintillator
transparency which is subject to aging.
To test the stability of the transparency, samples from four scintillator batches
have been taken from the acrylic tanks over time to study the transparency loss
of the scintillator in its experimental environment. Figure~\ref{stab} gives
the results of these repeated measurements. 
In all cases an initial decline stabilized, resulting in a loss rate of about 2 cm/day.
This rate of aging is acceptable for our experiment. The early samples BC1-4 (see fig.~\ref{al})
showed a much more rapid loss of transparency. It is interesting to note that
the aging process starts only after the blending. Test batches blended from concentrate
retains exhibit the same attenuation length as the original liquid sample. The aging
process is sped up by subjecting the liquid to elevated temperatures. In situ measurements
made for all scintillator batches by scanning the detector elements with a $\gamma$ source
confirm that the light attenuation has acceptable stability in time.

\section{Conclusion}

The formulation of a pseudocumene based 0.1$\%$ Gd loaded liquid scintillator
having good
long term stability and compatibility with acrylic has been reported. The  
measurements done during development and deployment of the scintillator have been
discussed.
The details of the scintillator preparation, which in our view were important 
for obtaining scintillator of high transparency and stability, were reported.

{\bf Acknowledgement}

The authors want to thank Prof. F. Boehm for his continued support and numerous
stimulating discussions. We also want to acknowledge the efforts of Dr. H. Hunter
who participated in the early development work and continued to help the project 
even after NE Technology Ltd. was no longer involved. The support 
of the Palo Verde
Collaboration and the active help of S. Beckman, Prof. J. Busenitz, Prof. G. Gratta and
Dr. J. Wolf was very much appreciated. Early development efforts were mainly driven 
by Prof. M. Chen and Dr. R. Hertenberger. Finally we acknowledge the excellent work of
the Bicron/SGIC laboratory technicians Michael Berkley and Tim Hill who were
responsible for the preparation of the Gd concentrate.

\end{document}